\documentclass[sigconf, nonacm]{acmart}

\usepackage{listings}

\AtBeginDocument{%
  \providecommand\BibTeX{{%
    \normalfont B\kern-0.5em{\scshape i\kern-0.25em b}\kern-0.8em\TeX}}}
    
\begin{document}

\title{Anti-patterns in Students' Conditional Statements}

% \author{Anonymous author}
% \email{anon@anon.anon}
% \affiliation{
%   \institution{Anonymous Institution}
%   \city{Anon}
%   \state{Anon} 
%   \country{Anon}
% }

% \author{Anonymous author}
% \email{anon@anon.anon}
% \affiliation{
%   \institution{Anonymous Institution}
%   \city{Anon}
%   \state{Anon}
%   \country{Anon}
% }

% \author{Anonymous author}
% \email{anon@anon.anon}
% \affiliation{
%   \institution{Anonymous Institution}
%   \city{Anon}
%   \state{Anon}
%   \country{Anon}
% }

\author{Etienne Naude}
\email{etienne.naude23.imperial.ac.uk}
\orcid{0000-0002-0010-3419}
\acmYear{2024}
\affiliation{
  \institution{Imperial College London}
  \city{London}
  \state{England}
  \country{United Kingdom}
  \postcode{SW7}
}

\author{Paul Denny}
\email{paul@cs.auckland.ac.nz}
\orcid{0000-0002-5150-9806}
\acmYear{2024}
\affiliation{
  \institution{The University of Auckland}
  \city{Auckland}
  \country{New Zealand}
  \postcode{1010}
}

\author{Andrew Luxton-Reilly}
\email{a.luxton-reilly@auckland.ac.nz}
\orcid{https://orcid.org/0000-0001-8269-2909}
\acmYear{2024}
\affiliation{
  \institution{The University of Auckland}
  \city{Auckland}
  \country{New Zealand}
  \postcode{1010}
}

\begin{abstract}

%Producing high-quality code is essential as it makes a code-base more maintainable. This allows software professionals to program more efficiently and use fewer resources on a given task. However, students learning to program are often given short, automatically marked programming tasks that do not need to be altered in the future, leading to poor-quality code containing many anti-patterns. We explore defects relating to conditional statements in students' code submissions to find that most students produced ``if/else return bool'', ``confusing else'', and ``nested if'' anti-patterns over the semester. Additionally, ``if/else return bool'' and ``confusing else'' combined to constitute nearly 60\% of the produced anti-patterns.

Producing high-quality code is essential as it makes a codebase more maintainable, reducing the cost and effort associated with a project.  However, students learning to program are often given short, automatically graded programming tasks that they do not need to alter or maintain in the future.  This can lead to poor-quality code that, although it may pass the test cases associated with the problem, contains \emph{anti-patterns} -- commonly occurring but ineffective or counterproductive programming patterns.  This study investigates anti-patterns relating to conditional statements in code submissions made by students in an introductory Python course. Our primary motivation is to understand the prevalence and types of anti-patterns that occur in novice code.  We analyzed 41,032 Python code submissions from 398 first-year students, using the open-source `qChecker' tool to identify 15 specific anti-patterns related to conditional statements. Our findings reveal that the most common anti-patterns are ``if/else return bool'', ``confusing else'', and ``nested if'', with ``if/else return bool'' and ``confusing else'' alone constituting nearly 60\% of the total anti-patterns observed. These anti-patterns were prevalent across various lab exercises, suggesting a need for targeted educational interventions. Our main contribution includes a detailed analysis of anti-patterns in student code, and recommendations for improving coding practices in computing education contexts.  The submissions we analyse were also collected prior to the emergence of generative AI tools, providing a snapshot of the issues present in student code before the availability of AI tool support.

\end{abstract}




\maketitle

\newpage
\section{Introduction and Background}
%Code style is generally linked to notions of quality. However, software quality has many facets. These are described in software product quality models~\cite{ Boehm:SWQUALITY76,Dromey:SWQUALITY96,ISOIEC:SWSTANDARD2501013,McCall:SWQUALITY77} and include characteristics such as efficiency, robustness and maintainability. The expectation is that architects and programmers address these by making appropriate design- and code-related decisions during development. Code style is one aspect of quality that impacts  \textit{maintainability}. Maintainability is a product quality characteristic that affects the ongoing costs of product development. The various quality models cite different sub-characteristics for maintainability, for example, some include testability as a sub-characteristic, but most include factors that relate to how easy it is to \textit{understand} and \textit{change} the software.

High-quality code has a wide range of characteristics, including being correct, efficient, robust, and being easy to maintain~~\cite{ Boehm:SWQUALITY76,Dromey:SWQUALITY96,ISOIEC:SWSTANDARD2501013,McCall:SWQUALITY77}.  We are most interested in \textit{maintainability} because it includes quality characteristics that relate to early programming concepts, such as readability and comprehensibility of code. These code quality characteristics are among the most frequently cited issues contributing to the reviewability and maintainability of code across a range of white and grey literature \cite{10.1145/3236024.3236080}. 
%Maintainable code helps professionals to program more efficiently and create fewer errors in their work, more efficiently and create fewer errors in their work, and as such, it is crucial for a software developer to produce high-quality code.  However, code quality is not entirely empirical. While there are aspects to code quality which can be measured empirically, for example, the code's efficiency or the amount of dead code, there are other aspects which are more challenging to measure empirically, such as using a consistent code style since programmers might value different aspects of code style more when evaluating code quality. 
To improve maintainability, automated tools such as ESlint\footnote{https://eslint.org
}, Pylint\footnote{https://pypi.org/project/pylint/} and PMD\footnote{https://pmd.github.io/} have been deployed to enforce coding standards and detect code smells \cite{10.1145/1953355.1953379} by focusing on anti-patterns.  Anti-patterns are common coding practices that may be ineffective or counterproductive and can be identified using static analysis.

Among the most common structures in code is the \texttt{if} statement appearing in about 35\% of novice students' Python code, and the \texttt{else if} statement occurring in about 20\% \cite{10.1145/3013499.3013500}. \texttt{If} statements, \texttt{else} statements and \texttt{elif} statements are simple structures that allow code to follow logical branches based on specific conditions. Although conditional logic is a crucial part of any programming, students often struggle with the logic of their programs \cite{10.1145/3373165.3373169,9463041}. 

For example, a typical exercise for students learning to write conditional statements is as follows:

\begin{quote}
\itshape
    Write a function named get\_last\_letter\_dictionary(sentence) 
    which takes a sentence as a parameter and returns a dictionary
    where the key is a letter and the value is a unique list of words
    that have the key letter as their last letter. You should convert
    the entire sentence to lowercase before processing.
\end{quote}

A typical solution produced by a CS1 student in a large introductory programming course at \textit{Anonymized for review} is reproduced below.

\begin{verbatim}
    def get_last_letter_dictionary(sentence1):
      sentence1 = sentence1.lower()
      dict1 = {}
      words = list(set(sentence1.split()))
      for i in words:
        if dict1.get(i[-1]):
          dict1[i[-1]].append(i)
        else:
          dict1[i[-1]] = []
          dict1[i[-1]].append(i)
      return dict1
\end{verbatim}

In this code, the line \texttt{dict1[i[-1]].append(i)} is repeated in both the \texttt{if} block and the \texttt{else} block, and therefore the conditional could be refactored to reduce repetition as follows:
%it is the only line of code inside the if block. As such, this line can be taken outside of the conditional code since it is not dependent on the condition. However, this would leave an empty if statement, so the condition can be inverted so that the logic for the else statement is run inside the if block; this leaves the conditional statement in the following form:

%In this code, several factors might lower the submission quality, making it slightly more challenging to read and maintain, such as converting. A straightforward example is the line "dict1[i[-1]].append(i)", which is repeated in both the if block and the else block, and it is the only line of code inside the if block. As such, this line can be taken outside of the conditional code since it is not dependent on the condition. However, this would leave an empty if statement, so the condition can be inverted so that the logic for the else statement is run inside the if block; this leaves the conditional statement in the following form:

\begin{verbatim}
    ...
        if not(dict1.get(i[-1])):
          dict1[i[-1]] = []
        dict1[i[-1]].append(i)
    ...
\end{verbatim}

In this work, we aim to understand the extent to which anti-patterns involving conditional statements are produced by students in CS1.   We answer the following research questions:  
%The duplicated code is an example of an anti-pattern that can be detected using static analysis tools.  %; without reading the context of the rest of the code, the duplicated code in the if/else block could be detected, and a suggestion to the programmer could be made. By detecting anti-patterns, we cannot suggest entirely new solutions to a task, but we can recommend minor adjustments.

%We have developed two research questions using anti-patterns to help us understand the code quality issues surrounding conditional statements.

\begin{itemize}
    \item[\textbf{RQ1}] What are the most common anti-patterns that first-year students produce while writing conditional statements?
    \item[\textbf{RQ2}] Do specific conditional anti-patterns occur more frequently in certain tasks than in other tasks?
\end{itemize}

\section{Related Work}
%==========

A range of prior work has been closely related to studying conditional statements, anti-patterns and code quality. \citet{9463041} studied students' comprehension of \texttt{if} statements and found that only 35\% of intermediate students understood that in conjoined \texttt{if}  statements (\texttt{if}  statements that use logical operators such as \texttt{AND}  to join two expressions), the first expression would be evaluated before the second expression. Failing to understand the order in which boolean expressions are evaluated could lead to logical errors.  For example, the code snippet:
\begin{verbatim}
    if(arr[0] == 2 AND len(arr) > 1):
\end{verbatim}
would throw an \texttt{out of range} error if the list \texttt{arr} is empty, while the following code will execute without error in the same circumstances:
%while the second snippet will not evaluate "arr[0] == 2" since the condition of the if statement fails at "len(arr) > 1".
\begin{verbatim}
    if(len(arr) > 1 AND arr[0] == 2):
\end{verbatim}

\citet{9463041} found that students' comprehension of code involving compound boolean expressions was greatly increased when the code was refactored to use individual if statements with simpler expressions. They found that 54\% of students recognised that the first \texttt{if} statement would be evaluated before the second and that the second might not be evaluated.

\citet{izu2022resource} explored a system to help students refactor code.  They provided novice students with tasks to refactor three short programs focused on conditional statements and compared the length of code written by these students to the cohort of students from the previous year who did not complete the refactoring tasks. All of the four tasks provided to students were, on average, written using more compact code by the students who completed the training on refactoring. However, as the authors discuss, while code length is an indicator of code quality, it is not a direct measure.

\citet{Edwards2019CanIS} used FindBugs to analyse students' Java code. FindBugs analyses both the correctness of code and generates warnings for code that exhibits quality issues (such as unused code). While FindBugs analyses 424 separate issues in code, only 76 issues were detected in the data set, and \citeauthor{Edwards2019CanIS} filtered these down to 55 issues which they deemed ``useful'' for analysing code quality rather than other aspects of code. Out of the 55 useful issues, four were related to \texttt{if} statements. \citeauthor{Edwards2019CanIS} found that 92\% of students encountered warnings on their work at some point during the course, while only 7\% of 149,054 task submissions contained these same warnings.

\citet{10.1145/3478431.3499310} aimed to replicate the work with FindBugs completed by \citeauthor{Edwards2019CanIS}; however, they analysed substantially more code submissions, targeting assignments which require multiple hundreds or thousands of lines of code to complete rather than a few lines. This study also involved students from CS1, CS2 and CS3 courses and included 255,222 submissions by 4,244 students. They found that 92\% of students encountered the same set of code issues as the original study.%, which means these two studies found a difference of 0.11\% of students encountering the set of issues, confirming the results of the original study.

%\citet{10.1145/3160489.3160500} examined a range of anti-patterns produced by students, counting how many students encountered each anti-pattern across four courses. Twelve out of the 16 anti-patterns examined were focused on conditional statements while the remaining four concentrated on variable assignment. When analysing the students' code, they found "if return condition" was found to be the most prevalent anti-pattern and was presented by 47.7\% of the students taking CS1 papers.

\citet{10.1145/3160489.3160500} examined a range of anti-patterns produced by students, counting how many students encountered each anti-pattern across four courses. Twelve out of the 16 anti-patterns examined were focused on conditional statements while the remaining four concentrated on variable assignment. Their analysis revealed that the "if return condition" (IFRC) anti-pattern was the most prevalent, observed in 47.7\% of the students taking CS1 papers. Additionally, they found that a significant portion of the students exhibited multiple anti-patterns, with half of the students submitting code containing two or more of these issues. This highlights the importance of addressing semantic style issues early in programming education, as even students in advanced stages of their studies, such as those in their fourth year of a highly competitive Software Engineering program, showed similar problems. This study underscores the value of providing formative feedback on semantic style to improve students' coding practices and overall code quality.

%\citet{10.1145/3059009.3059061} explored anti-patterns in students' Java code and how often students fixed specific anti-patterns in their code. \citeauthor{10.1145/3059009.3059061} found that students who were using tools built into their IDE to report these defects produced slightly more anti-patterns than those who did not. They also found that students often tidied up their code with 49.1\% of "EmptyIfStmt" anti-patterns fixed before the students' final submissions. 

\citet{10.1145/3059009.3059061} conducted a comprehensive study on the presence of code quality issues in novice programmers' Java code, focusing on anti-patterns. They analyzed a vast dataset from the Blackbox database, finding that students frequently encountered quality issues including related to program flow and choice of programming constructs. Specifically, issues such as ``EmptyIfStmt'' were common, with 49.1\% of these anti-patterns being fixed before the final submissions, indicating that students often tidied up their code. However, the overall rate of fixing issues was low, especially for more complex problems like modularization.
Interestingly, the study also found that students using integrated development environment (IDE) tools to report code defects produced slightly more anti-patterns than those who did not use such tools. This suggests that the mere presence of these tools may not be sufficient to improve code quality and that additional instructional support may be necessary to help students recognize and correct these issues.

\citet{rechtavckov2024catalog} recently published a comprehensive catalog of code quality defects commonly found in CS1 programs. They identified and categorized 80 distinct defects, including redundant if-else statements, providing a structured framework for understanding and addressing these issues. To determine the importance of these defects, they conducted a survey with 72 educators, focusing on the severity and prevalence of various defects. The survey revealed that defects involving conditional statements, such as redundant if-else constructs and unnecessary parentheses, were among those considered the most important by teachers. These defects are not only common but also significant in terms of their impact on code readability and maintainability.

%Similar to our work, 
%\citet{10.1145/3478431.3499415} explored a variety of anti-patterns using the term ``defects'', including many defects involving \texttt{if} statements which is also our focus. The most common anti-pattern found in conditional statements was ``redundant if-else''. This defect is broader than the anti-patterns detected by qChecker since it encapsulates "if/else return bool", "Duplicate If/Else Statement", "if return bool", "if/else assign return", "duplicate if/else body", "if/else assign bool" and, "if/else assign bool return". They found that "redundant if-else" was found in 3.8\% of submissions, while the next highest prevalence amongst the conditional statement defects averaged 1.8\%. 

Most similar to our work, \citet{10.1145/3478431.3499415} explored a variety of anti-patterns using the term ``defects,'' focusing on a broad range of issues, including those involving \texttt{if} statements, which align with our primary focus. They identified ``redundant if-else'' as the most prevalent anti-pattern for certain simple programming tasks. Their analysis of 114,000 solutions to 161 coding problems in Python revealed that most correct solutions contain some defects, and that ``redundant if-else'' was the most common conditional statement defect. 

%\citet{rechtavckov2024catalog} recently reported a catalogue of code quality defects common in CS1 programs and surveyed teachers about the importance of each defect category, reporting several defects involving conditional statements to be of high importance to teachers.

Much of the prior work exploring anti-patterns and cataloguing issues present in novice code has taken place prior to the emergence of generative AI, for which there has been a significant surge of interest in computing education \cite{denny2024computing, 10.1145/3657604.3662036, 10.1145/3626252.3630927, 10.1145/3623762.3633499, 10.1145/3626252.3630909}.  With the adoption of AI tools capable of generating code, it is possible that we may observe a reduction in the prevalence of anti-patterns, as the expectation is that these tools tend to produce code that adheres to best practice \cite{wadhwa2024core, lin2024llmbasedcodegenerationmeets}.  This certainly warrants investigation, and datasets like the one we have collected and analyzed in this work could be useful for exploring this.  The submissions in our dataset were produced before the availability of AI tools -- replicating this work using code written more recently could reveal important trends about the ways that students are adopting and using generative AI in computing courses.

%SUPER SIMILAR

\section{Method}

We analysed a range of final Python code submissions from first-year students at \{\emph{Institution Anonymized}\} enrolled in a CS2 course. This course 
%Computer Science 130. Computer Science 130 
is designed for students who have had some programming experience before but does not assume knowledge of a specific language.  The course, therefore, includes a recap of programming concepts using Python.  The course includes 22 laboratories over a 12-week semester (i.e., 2 laboratories per week for most weeks).  Each laboratory consists of a set of small exercises and are designed to take approximately two hours to complete.

Fourteen of the laboratories focused on writing code.  From these 14 laboratories, we removed any submissions that did not contain valid Python code. The total number of valid submissions across the 14 selected labs was 41,032, produced by 398 students.  The topics for each of the 14 labs are listed in Table \ref{tab:Labs}.

\subsection{qChecker static analyzer}
The Python package `qChecker' is a Python library designed to find statement-level anti-patterns in syntax trees and provide feedback to programming novices\footnote{https://github.com/James-Ansley/qchecker}. qChecker analyses Python programs for 25 different anti-patterns, 14 of which are focused on conditional statements. The anti-patterns which this library explores are based on those explored in works by \citet{10.1145/3160489.3160500} and \citet{10.1145/3478431.3499415} but splits some anti-patterns into smaller categories, and translates the relevant anti-patterns reported by \citet{10.1145/3160489.3160500} into Python.

\begin{table*}[]
    \centering
    \begin{tabular}{cll}
        \toprule
        \# & Topic & Description  \\
        \midrule
1 & Introduction & Focused on introducing students to Python and providing them with the basic tools to produce Python code. \\
2 & Modularity & Focused on splitting code up into readable structures and introducing students to functions. \\
3 & Testing &  Introduced basic testing and debugging practices. \\
4 & Exceptions & Helped students write resilient code by catching potential exceptions. \\
5 & Classes & Introduced students to classes in Python and explored some of the use cases of classes. \\
6 & OOP & Progressed the classes lab to develop the skills the students were learning with object-oriented programming. \\
7 & Recursion & The recursion tasks were meant to help students understand the difficulties in creating recursive code.\\
8 & Stacks & Was the first in a series about different data structures. \\
9 & Queues & Followed the series about different data structures. \\
10 & Lists &  Helped students prepare for the next labs by showing them how to create lists of objects.\\
11 & Linked Lists & Help students grasp linked lists' underlying structures and operations, which improve their efficiency.\\
12 & Hash tables & Introduced students to hash tables and showed students when to use them.\\
13 & Trees & Introduced the tree data structure and gave the students the base knowledge to work on the BST lab. \\
14 & BST & Followed on from the trees lab showing students how to create a binary search tree. \\
        \bottomrule
    \end{tabular}
    \caption{The topics of the lab assignments.}
    \label{tab:Labs}
\end{table*}

\subsection{Anti-patterns}

We analysed each of the 14 labs using qChecker. We filtered out the anti-patterns that did not relate to conditional statements to leave the 15 anti-patterns relating to conditional statements that the qChecker analysed. All of the conditional anti-patterns were present in the dataset. The specific anti-patterns qChecker examined regarding conditional statements have been listed in subsections \ref{anti-pattern-start} to \ref{anti-pattern-end}, with examples for each anti-pattern. %Our approach differs from the approach by \citet{10.1145/3478431.3499415} by using finer definitions for some of the explored anti-patterns and by performing the research using students who had some yet not extensive programming background.

\subsubsection{If Else Return Bool}\label{anti-pattern-start}

The same Boolean is returned in the if and the attached else statement. In this case, it could just return the Boolean.

\begin{verbatim}
    if(cond):
        return bool 
    else:
        return not(bool)
\end{verbatim}

\subsubsection{Confusing Else}

Detects nested if-else statements that do not need to be nested since the initial else statement has no other code other than the secondary if-else. This means the if-else statements can be flattened.

\begin{verbatim}
    if(cond):
        a +=1
    else:
        if(cond2):
            b +=1
        else:
            c += 1
\end{verbatim}
 
\subsubsection{Nested If}

Similarly to ``confusing else'', it detects nested if statements that do not need to be nested since the initial if statement has no other code besides the secondary if statement. This means the if-else statements can be flattened by merging the conditions into one if statement with "and".

\begin{verbatim}
    if(cond):
        if(cond2):
\end{verbatim}
 
\subsubsection{Duplicate If/Else Statements}

Duplicate code in the else and the if bodies. Since this code does not rely on the condition, it should be moved outside the if/else statement.

\begin{verbatim}
    if(cond):
        a +=1
        b +=1
    else:
        c += 1
        b += 1
\end{verbatim}
 
\subsubsection{If Return Bool}

The unnecessary use of an if statement since the condition of the if statement could be returned instead.

\begin{verbatim}
    if(cond):
        return bool 
    return not(bool)
\end{verbatim} 
 
\subsubsection{Empty If Body}

The body of an if statement contains no functional code, such as a variable being self-assigned or the use of ``pass''.

\begin{verbatim}
    if(cond):
        pass
\end{verbatim}

\subsubsection{Unnecessary Elif}

An elif uses the inverse of the condition to the if statements condition. This means the elif can be replaced with an else.

\begin{verbatim}
    if(cond):
        cond += 1
    elif(not(cond))
        print(cond)
\end{verbatim} 
    
\subsubsection{Else If}

An if statement inside an otherwise empty else statement, in which case an elif statement could replace it.

\begin{verbatim}
    if(cond):
        cond += 1
    else:
        if(not(cond)):
            print(cond)
\end{verbatim}     
    
\subsubsection{Empty Else Body}

Similar to the ``empty if body'' anti-pattern. It detects whether the body of an else statement contains no functional code, such as a variable being self-assigned or the use of ``pass''.

\begin{verbatim}
    if(cond):
        cond += 1
    else:
        pass
\end{verbatim}

\subsubsection{Unnecessary Else}

Similar to ``duplicate if/else statements'', although it differs since the entire else statement can be removed due to duplicate code, whereas the "Duplicate If/Else Statements" anti-pattern would still require the else statement since it contains other functional code. 

\begin{verbatim}
    if(cond):
        a +=1
        b +=1
    else:
        b += 1
\end{verbatim}

\subsubsection{Sev. Duplicate If/Else Statements}

Similar to the ``duplicate if/else statements'' but detects several duplicate statements.

\begin{verbatim}
    if(cond):
        a += 1
        
        b += 1
        print(b)
    else:
        c += 1
        
        b += 1
        print(b)
\end{verbatim}

\subsubsection{If/Else Assign Return}

An if/else statement that assigns a variable and then immediately returns the newly assigned variable.

\begin{verbatim}
    if(cond):
        name = a 
    else:
        name = b
    return name
\end{verbatim}

\subsubsection{Duplicate If/Else Body}

The entire body of the if statement is duplicated into the else statement.

\begin{verbatim}
    if(cond):
        b +=1
    else:
        b += 1
\end{verbatim}

\subsubsection{If/Else Assign Bool}

A Boolean could be assigned to the value of the condition of an if statement rather than forming an if statement.

\begin{verbatim}
    if(cond)
        name = bool
    else
        name = not(bool)
\end{verbatim}

\subsubsection{If/Else Assign Bool Return}\label{anti-pattern-end}

– Similar to ``if/else assign return'' and ``if/else assign bool'', however, the critical difference is assigning a boolean and returning it. 

\begin{verbatim}
    if(cond):
        name = bool
    else:
        name = not(bool)
    return name
\end{verbatim}

\begin{figure*}[htp]
    \centering
    \includegraphics[width=\textwidth]{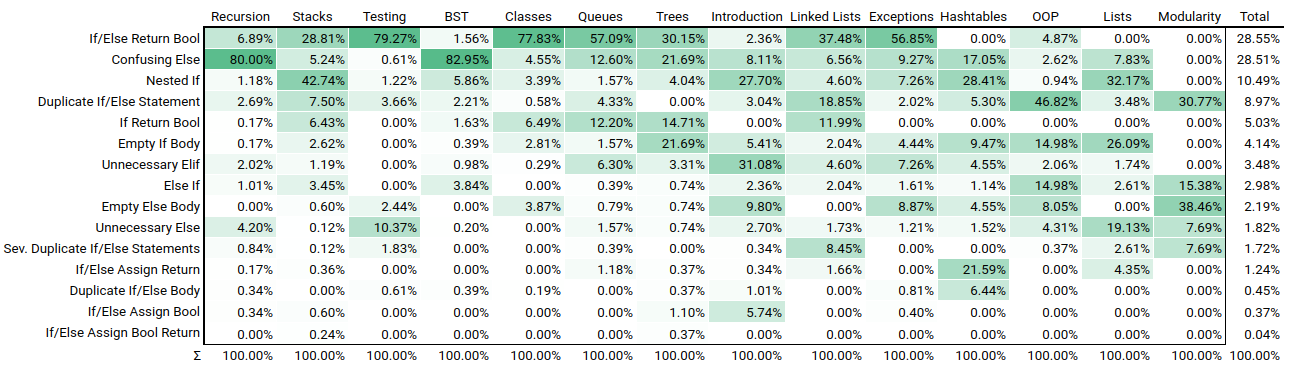}
    \caption{Prevalence of each anti-pattern.}
    \label{fig:percent}
\end{figure*}

\begin{figure*}[htp]
    \centering
    \includegraphics[width=\textwidth]{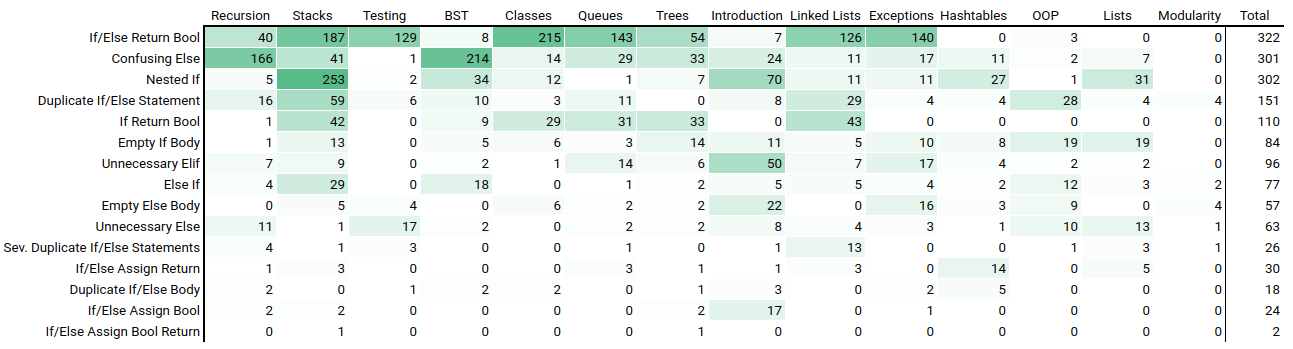}
    \caption{Unique students who encountered the anti-pattern.}
    \label{fig:students}
\end{figure*}

\section{Results}

\begin{figure}[htp]
    \centering
    \includegraphics[scale=0.55]{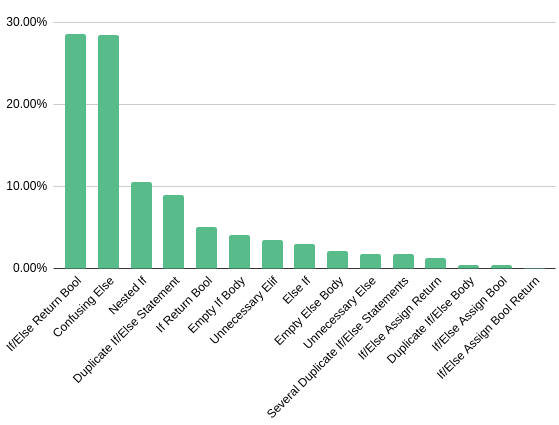}
    \caption{The relative proportions of conditional anti-patterns.}
    \label{fig:graph}
\end{figure}

We used qChecker to evaluate all the lab submissions and report the results in Figure \ref{fig:percent} and Figure \ref{fig:students}.  Figure \ref{fig:percent} shows the percentage of prevalence of each anti-pattern for all questions in a lab. For example, in the recursion lab, 80\% of the encountered anti-patterns were ``confusing else'' anti-patterns. As such, this percentage is reported in proportion to the other anti-patterns. The last column of this table shows the overall prevalence of each anti-pattern across all the different labs combined. The total proportions are also represented in Figure \ref{fig:graph}, which shows a bar graph displaying the sharp drop off of the prevalence of the anti-patterns.

Figure \ref{fig:students} shows the number of students who encountered each issue in a given lab, with the final column showing the total number of unique students (out of 398) who encountered each issue across all of the labs. The columns in Figure \ref{fig:percent} and Figure \ref{fig:students} were sorted in descending order of labs containing the most anti-patterns per line of code, and the rows were sorted in descending order of the most common anti-patterns per line of code.

We detected a total of 7,491 anti-patterns across the 41,032 submissions. Each anti-pattern was produced by an average of 14.4 students per lab (standard deviation of 37.0). We found that the most common anti-patterns detected were ``if/else return bool'' and ``confusing else'', which occurred roughly the same number of times and combined to constitute 57.1\% of the total anti-patterns detected. However, ``if/else return bool'' was produced by 7.00\% more students. The anti-pattern ``if/else return bool'' was also produced by enough students to be above the second standard deviation (88.4 students) in six of the labs, while ``confusing else'' was produced above this in two of the labs and "nested if" was seen above the second standard deviation once. The only lab with multiple data points above the second standard deviation was the ``stacks'' lab. There were no anti-patterns that occurred in every lab, and there were no labs that contained every anti-pattern.

\section{Limitations}

%This work has some limitations.  There could be other anti-patterns involving conditional statements that qChecker does not detect. %Students might have also received different levels of help in specific labs since they might be more inclined to ask for help later in the semester once they have found the support channels or near the start of the semester when they have fewer assignments due and can spend more time focusing on the lab tasks. 
%Students might have shared or copied code that could have an anti-pattern in it and, through a network of group messages, could propagate one instance of a solution to many students.
%Additionally, the study was conducted with students from one course at one university in one semester. These results might differ in other teaching contexts. % the research was conducted using different data from other universities or with students with differing levels of skill to those in this semester of the paper. 
%Additionally, the results might vary based on the programming languages the students use to complete the exercises. We also acknowledge that anti-patterns are a measure of one aspect of code quality and there are many other methods to measure code quality.

This work has several limitations that should be considered when interpreting the results. Firstly, we rely on the qChecker tool for detecting anti-patterns. While qChecker is effective at identifying a range of common issues, it may not detect all possible anti-patterns, particularly those that are more context-specific or less well-defined. Additionally, qChecker focuses on specific anti-patterns related to conditional statements, so other types of code quality issues may not have been captured in this study.

The study was also conducted with a single cohort of students from one university during one semester. This limits the generalizability of the findings, as the prevalence and types of anti-patterns observed may differ in other educational contexts, with different student populations, curricula, or programming languages.  

Finally, given the current availability of generative AI tools that can be used to generate code, replicating this work now may reveal an impact that these tools have had on the occurrences of anti-patterns.  The dataset we use comprises submissions made before the widespread adoption of generative AI tools for coding, and so may not fully reflect the current state of student programming practices.

\section{Discussion}

The laboratory exercises that were used in our study are likely to have influenced our results.  Certain exercises are likely to require solutions that use specific structures, and these structures may be prone to specific anti-patterns.  For example, tasks involving binary search trees often contain an if-else conditional statement to perform actions on the left or right sides of the tree. This may result in the ``confusing else'' anti-pattern being more commonly observed in this task because the use of many if-else statements provides more opportunity to produce the ``confusing else'' anti-pattern.

Our finding that the most common anti-patterns produced by students were ``if/else return bool'' and ``confusing else'' is consistent with the previous work produced by \citet{10.1145/3478431.3499415} and \citet{10.1145/3160489.3160500}. %The findings of \citeauthor{10.1145/3478431.3499415} that "redundant if-else" was the most common anti-statement that was detected supports our results since the "redundant if-else" anti-pattern encapsulates the "if/else return bool" anti-pattern and \citeauthor{10.1145/3478431.3499415} did not detect "confusing else" anti-patterns. 
Our work is also consistent with the findings by \citet{9463041} that students lack understanding of compound if statements. Students who are unsure about the order of operations in  compound if statements may choose to nest statements instead, causing the occurrence of ``confusing else'' and ``nested if'' anti-patterns.

As with work by \citet{Edwards2019CanIS} and \citet{10.1145/3478431.3499310}, we found that many students face quality issues during the semester. We encourage teachers to address common anti-patterns in student code explicitly.  Recent work by \citet{kirk_et_al_2024} proposes a set of principles to teach code style to students who are learning to program.  It would be interesting to map the specific anti-patterns produced by students to the principles and explore the impact of using student-generated examples to illustrate the principles of good style when teaching.

We also encourage teachers using automated assessment tools to grade the correctness of student work to incorporate a system such as qChecker into their autograding pipeline in order to raise student awareness of the quality issues in their code. %As \citet{Birillo2022HyperstyleAT} analysed the code quality of students with code using Java, using qChecker to provide similar feedback to students using Python would be interesting. By using qChecker to provide feedback to students, its warnings might aid students in becoming more productive and help them fix the specific reoccurring quality issues they are facing.

Another potential method to promote high-quality code would be to use a system similar to the work presented by \citet{10.1145/3408877.3432526} and \citet{izu2022resource} and present exercises to students with poor quality code --- asking them to refactor it into higher quality code. Such exercises could leverage the findings in this paper by targeting the anti-patterns that we found to be the most common. %Further research needs to be done to see if exercises like this would reduce the number of anti-patterns in future exercises. 

There are many further aspects of student code quality that relate to conditional statement anti-patterns that could form the basis for future research. One potential research area could be whether these issues continue with the same prevalence as students progress in their studies.
Our study focused on identifying anti-patterns in a first-year course, but it would be valuable to investigate whether these issues persist as students progress to more advanced courses, and eventually into professional environments.  Tracking students over time could also provide insights into the effectiveness of any interventions that were introduced to address anti-patterns, such as the automated feedback suggestions presented earlier in this section. 

%Finally, exploring other anti-patterns not relating to conditional statements could also provide valuable insight into how students produce code. 

%Another interesting area of future research could be whether there is a strong correlation between the students' grades and the quality rating of their code. Since the short form assignments we examined are marked automatically, and their code quality is not a metric that is evaluated, the quality of the code produced might not be an essential aspect of these assignments. As such, tasking students with longer-form assignments which they would need to work on and contribute to for multiple weeks in a row might be a method to incentivise students to produce high-quality code so that it is easier for them to add to their code in the future. Since some students might work on large assignments only shortly before the deadline, providing subsections of a larger assignment to them each week might help them produce additional work, provide a better environment to produce high-quality code and prepare students for industry work in which they would need to build large scale applications. Each week's assignments could be broken down into simple tasks, each of which requires only a few lines of code, similar to the current lab structure. However, the difference would be that each task builds on the previous tasks. One other benefit of structuring assignments like this is that students would have completed a small scale project by the end, which could be added to their portfolio, impacting their employability.  

Finally, we recognize that the rise of generative AI will likely impact the quality of code that students read and write.  The data reported in this study is derived from student submissions to programming tasks that were collected prior to the release of commonly used generative AI tools such as Copilot and ChatGPT \cite{denny2023conversing, ouh2023chatgpt}.  It would be interesting to replicate this work in the near future to determine the extent to which such tools has impacted the anti-patterns present in student-submitted code.

\section{Conclusion}

In this work, we explore the anti-patterns that are present in code submitted by students in a first-year Python programming course.  Understanding the prevalence and types of anti-patterns in novice programmers' code is useful for guiding the development of targeted educational interventions to improve coding practices early in the learning process.  We analyzed 41,032 Python code submissions from 398 first-year students to investigate the occurrence of anti-patterns in conditional statements. Using the open-source tool `qChecker', we identified 15 specific anti-patterns related to conditional statements. The key findings from our analysis revealed that the most common anti-patterns are ``if/else return bool'', ``confusing else'', and ``nested if'', with the first two anti-patterns alone accounting for nearly 60\% of all identified issues. These patterns were consistently observed across various lab exercises, indicating a widespread problem that warrants attention.  We recommend the integration of anti-pattern detection tools, like qChecker, into automated grading systems, as this may help to raise student awareness about code quality issues.  We also suggest that educational interventions such as targeted exercises may be useful for reducing the prevalence of these anti-patterns.  Finally, with the rise of generative AI tools for coding, future research should investigate how these tools influence the occurrence of anti-patterns in student submissions.

%From our results, we were able to draw conclusions about both of our research questions. To RQ1 (What are the most common anti-patterns which first-year students produce while writing conditional statements?) 
%The most common anti-patterns students produced were "If/Else Return Bool" and "Confusing Else". These two anti-patterns had a high prevalence and constituted 28.55\% and 28.51\%, respectively, of the anti-patterns produced. "If/Else Return Bool" was also the most commonly produced anti-pattern in six of the 14 labs.  We also found that the prevalence of anti-patterns was not evenly distributed across the different laboratories.  Most laboratories have one or two issues that are prevalent rather than all those being prevalent for a lab. This suggests that specific tasks and topics tend to be answered in ways in which certain anti-patterns are more likely to occur.

%NEED A FINAL SENTENCE THAT WRAPS UP THE CONTRIBUTION OF THE WORK

\bibliographystyle{ACM-Reference-Format}
\bibliography{references}

\end{document}